\documentclass[aps,twocolumn,groupedaddress]{revtex4}
\usepackage{graphicx}%

\begin{document}
\bibliographystyle{apsrev}


\title[Lattice from 2 tilts]{Lattice parameters from direct-space images at two tilts}


\author{W. Qin}
\altaffiliation{Motorola PMCL/DDL/SPS, MD M360, Mesa AZ 85202}
\author{P. Fraundorf}
\email[]{pfraundorf@umsl.edu}
\affiliation{Physics \& Astronomy and Center for Molecular Electronics, U. of Missouri-StL (63121), St. Louis MO USA}


\date{\today}

\begin{abstract}
Lattices in three dimensions are oft studied from the ``reciprocal space'' 
perspective of diffraction. Today, the full lattice of 
a crystal can often be inferred from direct-space information about
three sets of non-parallel lattice planes. Such data can come 
from electron-phase or Z contrast images taken at two tilts, provided that 
one image shows two
non-parallel lattice periodicities, and the other shows a periodicity not
coplanar with the first two. We outline here protocols for measuring the 3D
parameters of cubic lattice types in this way. For randomly-oriented
nanocrystals with cell side greater than twice the continuous transfer
limit, orthogonal $\pm 15^{\circ }$ and $\pm 10^{\circ }$ tilt ranges might
allow one to measure 3D parameters of all such lattice types in a specimen 
from only two well-chosen images. The strategy is illustrated by 
measuring the lattice parameters of a $10$ nm WC$_{1-x}$ crystal in a 
plasma-enhanced chemical-vapor deposited thin film.
\end{abstract}

\maketitle

\section{Introduction}

If 10 nm crystals presented themselves to our perceptions as 10 cm 
hand specimens, new rules for direct-space crystallography might 
have emerged.  For example, if by tilting the specimen 
one could locate a set of 4-fold cross-fringes, from which a tilt 
(at $45^{\circ }$ to those fringes) by 
$35.3^{\circ }$ {\em just} brings one to a new set of fringes $15.5$ 
percent larger in spacing, then the hand specimen is likely face-centered-cubic.  
It's reciprocal lattice definitely includes a body-centered-cubic 
array like that characteristic of face centered-cubic-crystals.

Here we discuss the geometry of crystals from the perspective of lattice
imaging in direct space. The required instrumentation consists of a TEM able
to deliver phase or Z contrast lattice images of desired periodicities (e.g.
spacings down to half the unit cell side for cubic crystals), and a specimen
stage with adequate tilt (e.g. two-axes with a combined tilt range of $\pm
18^{\circ }$). For crystals with lattice spacings of $0.25$nm and larger,
many analytical TEM's will work, while a high resolution microscope with
continuous contrast transfer to spatial frequencies beyond $1/{(0.2}${nm}${)}
$ can do this for most crystals. We demonstrate the process experimentally
by determining the lattice parameters of a tungsten carbide nano-crystal
using a Philips EM430ST TEM. Appropriately orienting the crystal, so as to
reveal its three-dimensional structure in images, is a key part of the
experimental design and will be discussed in detail. In the process,
strategies for supporting on-line electron-crystallography, for
three-dimensional lattice-correlation darkfield studies of nanocrystalline
and paracrystalline materials, and for stereo-diffraction analyses, are
suggested as well.

\section{Calculations}

\subsection{Overview of a ``tilt protocol'' in action}

For the stereo lattice-imaging strategy discussed here, low Miller index
(hence large) lattice spacings are both easier to see, and more diagnostic
of the lattice. Orientation changes directed toward the detection of such
spacings are needed. In this section, tilt protocols optimized for getting
3D data from one abundant class of lattice types (namely cubic crystals) are
surveyed. We begin with a list of candidate lattice types (e.g. f.c.c. or 
h.c.p.) based on prior compositional, diffraction, or imaging data.

Three non-coplanar reciprocal lattice vectors seen along 2 different zone
axes are sufficient for inferring a subset of the 3D reciprocal lattice of a
single crystal. Often these are adequate to infer the whole reciprocal
lattice. Hence the goal of our experimental design is to look for at least 3
non-coplanar lattice spatial frequencies, in two or more images. We prefer
images with ``aberration limits'' $r_{a}$ smaller than the analyzed
spacings, to lessen chances of missing other comparable (or larger) spacings
in the exit-surface wavefield. So as to tilt from one zone to another, the
crystal must also be oriented so that the desired beam orientations are
accessible within the tilt limits of the microscope.


To illustrate, we exploit the fact (considered more fully below) that {\em %
all} cubic crystals will provide data on their three dimensional lattice
parameters if imaged down selected zones separated by $35.3^{\circ }$,
provided that spacings at least as large as half the cell side $a$ are
reported in the images.  With the images discussed here we
expect to ``cast a net'' for 3D data on any cubic crystals whose cell
side $a$ is larger than $2\times 0.19nm=0.38nm$.  More than 85\% of the cubic
close packed crystals and nearly 40\% of the elemental b.c.c. crystals
tabulated in Wyckoff \cite{Wyckoff}, for example, meet this requirement, as
of course would most cubic crystals with asymmetric units comprised of more
than one atom.

Although $35.3^{\circ }$ is too far for the eucentric tilt axis in our
microscope, combining two tilts gives us a range of $35.6^{\circ }$. Two
images therefore can be taken at orientations $35.3^{\circ }$ apart, namely
at ($\vartheta _{1}=15.0^{\circ }$, $\vartheta _{2}=9.7^{\circ }$) and ($%
\vartheta _{1}=-15.0^{\circ }$, $\vartheta _{2}=-9.7^{\circ }$),
respectively, where $\vartheta _{1}$ and $\vartheta _{2}$ are goniometer
readings on our double tilt holder. This yields an ``effective'' tilt axis
that runs perpendicular to the electron beam. Its azimuth is $123.5^{\circ }$
in the $xy$ plane of our images. The coordinate system used will be
discussed in more detail later. For the special case of fcc crystals, the
zones of nterest are [$001$] and [$112$]. Since the ($2\overline{2}0$)
lattice planes are parallel to both desired zones, the tilt must be along
these planes. That is, ($200$) fringes seen down a 4-fold symmetric [$001$]
zone must make an angle approaching $45^{\circ }$ with respect to the
effective tilt axis. Nearly a third of the randomly-oriented crystals
showing [$001$]-zone fringes will be sufficiently close \cite{QinThesis}.

The experimental results were unambiguous. We found many four-fold symmetric
images having spacings consistent with WC$_{1-x}$. This has an f.c.c.
lattice with $a=0.4248$nm \cite{Krai28,JCPD29}. When such a zone was
found with fringes making $45^{\circ }$ to the effective tilt axis in the
first image, a new spacing was seen in the second image making a 3D lattice
parameter measurement possible. The experiment is illustrated in Figure \ref
{Fig1}.

\subsection{Experimental designs}

Here we seek three non-coplanar periodicities from two images, although the
analysis also works if they are discovered singly in three images. Given
Miller indices $(h_{1}k_{1}l_{1})$ and $(h_{2}k_{2}l_{2})$ for any two
periodicities (i.e. vectors ${\bf g}_{1}$ and ${\bf g}_{2}$, respectively,
in the reciprocal lattice) of any crystal, first find zone indices $%
[u_{A}v_{A}w_{A}]$ of the beam direction ${\bf r}_{A}$ $\equiv {\bf g}%
_{1}\times {\bf g}_{2}$ needed to view both spacings in one image. The axis
for the smallest tilt that will make the beam orthogonal to a third
periodicity with indices $(h_{3}k_{3}l_{3})$, and reciprocal lattice vector $%
{\bf g}_{3}$, may then be defined by the vector ${\bf v}_{t}\equiv {\bf g}%
_{3}\times {\bf r}_{A}$. Lastly, zone indices $[u_{B}v_{B}w_{B}]$ for the
beam after the specimen has been tilted around this axis so as to image the
third periodicity, may be obtained from the expression ${\bf r}_{B}\equiv 
{\bf v}_{t}\times {\bf g}_{3}$. Note here that we treat the Bragg angle for
electrons as small (i.e. less than one degree). Thus the actual tilt
required will be a fraction of a degree less.

Although these cross product calculations can be done by first converting
for example to ``c-axis'' cartesian coordinates \cite{Phil4}, the simplest
determination of needed parameters is perhaps done using the metric matrix $%
G $ of a prospective lattice \cite{Boisen}: 
\begin{equation}
G\equiv \left[ 
\begin{array}{ccc}
{\bf a}\bullet {\bf a} & {\bf a}\bullet {\bf b} & {\bf a}\bullet {\bf c} \\ 
{\bf b}\bullet {\bf a} & {\bf b}\bullet {\bf b} & {\bf b}\bullet {\bf c} \\ 
{\bf c}\bullet {\bf a} & {\bf c}\bullet {\bf b} & {\bf c}\bullet {\bf c}
\end{array}
\right] \text{.}  \label{metric_equation}
\end{equation}
If we denote row vectors formed from Miller (or lattice) indices as $\langle
ijk|$, and column vectors as $|ijk\rangle $, then the zone $A$ indices obey $%
{\bf g}_{1}\bullet {\bf r}_{A}=\langle
h_{1}k_{1}l_{1}|u_{A}v_{A}w_{A}\rangle =0$ and ${\bf g}_{2}\bullet {\bf r}%
_{A}=\langle h_{2}k_{2}l_{2}|u_{A}v_{A}w_{A}\rangle =0$. From these two
equations, $[u_{A}v_{A}w_{A}]$ follows except for a multiplicative constant
which is not important. Similarly, the (possibly irrational) Miller indices
of the tilt axis $(h_{t},k_{t},l_{t})$ may be determined from ${\bf g}%
_{t}\bullet {\bf r}_{A}=\langle h_{t}k_{t}l_{t}|u_{A}v_{A}w_{A}\rangle =0$
and ${\bf v}_{t}\bullet {\bf g}_{3}=0=\langle
h_{t}k_{t}l_{t}|G^{-1}|h_{3}k_{3}l_{3}\rangle $ \cite{SpenceZuo}. Only in
this fourth equality does $G$ affect the calculation, and for cubic crystals
it then simply offers a multiplying constant. Finally, the zone $B$ indices
follow (to within a factor) simply from ${\bf g}_{t}\bullet {\bf r}%
_{B}=\langle h_{t}k_{t}l_{t}|u_{B}v_{B}w_{B}\rangle =0$ and ${\bf g}%
_{3}\bullet {\bf r}_{B}=\langle h_{3}k_{3}l_{3}|u_{B}v_{B}w_{B}\rangle =0$.

\begin{figure}
\includegraphics{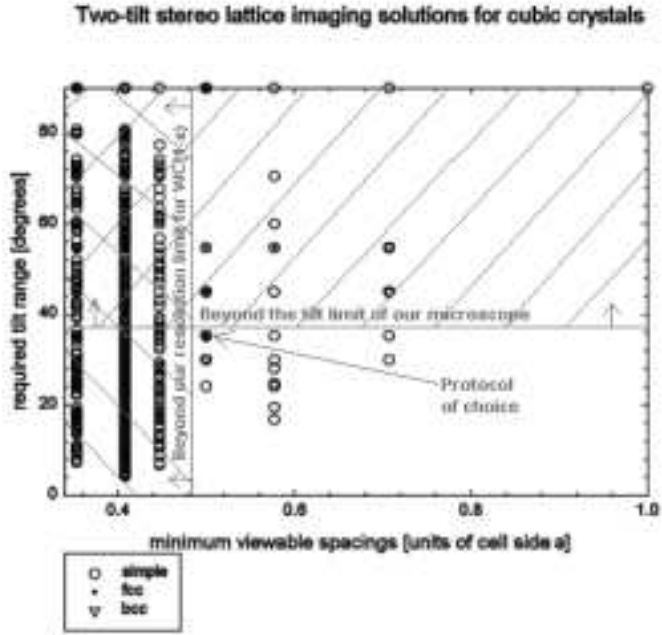}%
\caption{Tilt protocols for determining the lattice parameters of
face-centered, body-centered, and simple cubic crystals from two high
resolution images, plotted according to the required tilt range, and
required resolution limit in units of the cubic cell side. Superimposed on
this plot are the tilt range, and resolution in units of the unit cell side $%
a=0.4248$nm of WC$_{1-x}$, required for the experiment reported here.}
\label{Fig0}
\end{figure}

Two parameters which determine the attractiveness and feasibility of a given
experiment are the spatial resolution, and range of specimen tilts, that the
microscope is able to provide. For a given lattice type, it is useful to:
(i) go through the list of all pairs of periodicities calculating the tilt
between the zone associated with that pair and any third spacing of possible
interest, and (ii) rank the findings according to the minimum-spacing that
must be resolved, and the range-of-tilt that the specimen undergoes. This
has been done for face-centered, body-centered, and simple-cubic lattices,
and the results illustrated in Figure \ref{Fig0}, with examples for fcc and
bcc spacings no less than half of the cell side illustrated in Figure \ref
{BigFig}. Note that this calculation needs to be done only once for each
unit cell shape. Factors like the multiplicity of a given zone type might
also figure into the design of experiments with randomly-oriented crystals,
although we have not considered them here.

High tilts can be used to lower measured spacing uncertainties, in
directions perpendicular to the electron microscope specimen plane \cite
{Phil27}. Hence the protocols of interest for a given experiment may be those
which approach goniometer tilt limits, or at least the limits of specimen
tiltability.

Concerning the resolution limit to use, we suggest the spacing associated
with the end of the first transfer function passband {\em in the micrograph
of interest}, sometimes inferrable from regions in the image showing
disordered material. Even in this case, possible thickness and
misorientation effects warrant caution \cite{Spen36,Malm39}.

One may of course explore spatial frequencies in the specimen up to the
microscope information limit. However, spherical aberration zeros in the
transfer function introduce the possibility that the microscope will
suppress some spatial frequencies present in the subspecimen electron
wavefield. To lessen this problem, HREM images taken at different focus
settings could be compared, if the foci were chosen so that spatial
frequencies lost at one defocus are likely recorded at another. This would
allow measurement of three non-coplanar reciprocal lattice vectors, without
missing any whose (reciprocal) length is shorter than the longest among
those three.

Lastly, of course, the protocol chosen may depend also on the specimen. For
an image field containing hundreds of non-overlapping but randomly-oriented
nano-crystals, only two micrographs could allow one to measure the
three-dimensional lattice parameters of all cubic crystal types present with
cell sides $a$ greater than $2r_{a}$. On the other hand, for a single
crystal specimen of unknown structure, both a great deal of tilt range, and
considerable trial and error tilting (or guess work based on lattice models)
might be required before a single set of three indexable non-coplanar
spacings is found.

\subsection{Inferring 3D reciprocal lattice vectors from micrographs}

Consider a specimen stage with two orthogonal tilt axes (and associated
rotation matrices) $T_{1}$ and $T_{2}$, both perpendicular to the electron
beam (the second only so when the first axis is set at zero tilt). When the
specimen is untilted ($\vartheta _{1}=0^{\circ }$, $\vartheta _{2}=0^{\circ
} $), vectors in the reciprocal lattice of the specimen may be described, in
coordinates referenced to the microscope, as column vectors $|{\bf g\rangle }
$. When this reciprocal lattice vector is tilted to intersect the Ewald
sphere by some double tilt in the sequence $T_{2}(\vartheta _{2})$ then $%
T_{1}(\vartheta _{1})$, it's presence may be inferred from diffraction
patterns or micrograph power spectra. In our fixed coordinate system, $%
|g\rangle $ has become $|g_{m}\rangle $, where the $m$ means that $%
|g_{m}\rangle $ is associated with the lattice periodicity $d_{m}=1/{g_{m}}$
recorded on a micrograph. Using matrix notation, we might then write:

\begin{equation}
|{\bf g}_{m}\rangle =T_{1}(\vartheta _{1})T_{2}(\vartheta _{2})|{\bf %
g\rangle }  \label{gsubm}
\end{equation}
Components of $g_{m}$ may be determined from the polar coordinates ($g$, $%
\varphi $) of a spot in the power spectrum of a recorded image, following:

\begin{equation}
|{\bf g}_{m}\rangle \equiv \left( 
\begin{array}{c}
g_{mx} \\ 
g_{my} \\ 
g_{mz}
\end{array}
\right) =g\left( 
\begin{array}{c}
\cos (\varphi ) \\ 
\sin (\varphi ) \\ 
0
\end{array}
\right)  \label{gsubmx}
\end{equation}
where $g$ is the length of the diffraction vector (e.g. in reciprocal nm)
and $\varphi $ is its azimuth corrected for lens rotation.

Hence we can calculate the ``untilted-coordinates'' $|g\rangle $, of
reciprocal lattice objects at $|g_m\rangle $ inferred experimentally from
micrographs, using

\begin{equation}
|g\rangle =T_2^{-1}(\vartheta _2)T_1^{-1}(\vartheta _1)|{\bf g}_m\rangle
=A(\vartheta _1,\vartheta _2)|{\bf g}_m\rangle \text{,}  \label{g_eqn}
\end{equation}
where we've defined:

\begin{equation}
A(\vartheta _1,\vartheta _2)\equiv T_2^{-1}(\vartheta _2)T_1^{-1}(\vartheta
_1) \text{.}  \label{t2stuff}
\end{equation}

The resulting $xyz$ coordinates of reciprocal lattice features $g$,
associated with the crystal while in the untilted goniometer specimen
orientation, but referenceable from micrographs taken at any orientation,
provide the language we use for speaking of our measurements in three
dimensions.

\subsection{Calculating lattice parameters}

Given 3D cartesian coordinates of ``points'' in the reciprocal lattice of a
crystal, we are in much the same situation as if we had diffraction patterns
of the crystal from two directions containing three (or more) non-coplanar
spots. Hence methods for stereo-analysis of diffraction data \cite
{Phil4,Phil7} can be used at this point. We summarize in this context briefly. Given
measured reciprocal lattice vector coordinates, a natural next step is to
infer a basis triplet for the crystal's reciprocal lattice. Three alternate
paths to this basis triplet might be referred to as ``matching'',
``building'' \cite{Phil7}, and ``presumed'' \cite{Phil4}.

Given an experimental basis triplet from any of these sources, lattice
parameters ($a$, $b$, $c$, $\alpha $, $\beta $, $\gamma $), goniometer
settings for other zones, and many other things follow simply from the
oriented triplet matrix defined below:

\begin{equation}
{\bf W} \equiv \left[ 
\begin{array}{ccc}
a_x & a_y & a_z \\ 
b_x & b_y & b_z \\ 
c_x & c_y & c_z
\end{array}
\right] =\left[ 
\begin{array}{ccc}
a_x^{*} & b_x^{*} & c_x^{*} \\ 
a_y^{*} & b_y^{*} & c_y^{*} \\ 
a_z^{*} & b_z^{*} & c_z^{*}
\end{array}
\right] ^{-1}  \label{conversion}
\end{equation}

Given $W$, for example, cartesian coordinates {\em in the microscope} for
any direct lattice vector with indices $[uvw]$ follow from $|r\rangle
=W|uvw\rangle $, while cartesian coordinates for the reciprocal lattice
vector with indices $(hkl)$ may be predicted from $\langle g|=\langle
hkl|W^{-1}$. These rules of course include instructions for calculating
basis vectors of the lattice, such as ${\bf a}\equiv \lbrack 100]$, and
reciprocal lattice, such as ${\bf b}^{\ast }\equiv (010)$, and the angles
between. Moreover, the oriented cartesian triplet $W$ is simply related to
the metric matrix for the lattice in equation \ref{metric_equation} by $%
G=WW^{T}$. From $G$, of course, all the familiar orientation-independent
properties of the lattice follow \cite{SpenceZuo}, including cell 
volume $V_{cell}=\sqrt{%
\left| G\right| }$, Miller/lattice vector dot products $\langle
g_{hkl}|r_{uvw}\rangle =\langle hkl|uvw\rangle $, reciprocal lattice vector
and interplanar spacing magnitudes $g_{hkl}^{2}=1/d_{hkl}^{2}=\langle
hkl|G^{-1}|hkl\rangle $, lattice vector magnitudes $r_{uvw}^{2}=\langle
uvw|G|uvw\rangle $, reciprocal lattice dot products $\langle
g_{1}|g_{2}\rangle =\langle h_{1}k_{1}l_{1}|G^{-1}|h_{2}k_{2}l_{2}\rangle $,
interspot angles $\theta _{12}=\cos ^{-1}\left[ \langle g_{1}|g_{2}\rangle
/(g_{1}g_{2})\right] $, etc.

Even before a basis triplet is selected, indexing of observed reciprocal
lattice vectors $g$ can be attempted by matching spacings and interspot
angles to candidate lattices. Because of the uniqueness of non-coplanar
triplets in three dimensions (providing at least a significant subset of the
whole reciprocal lattice), the matches are very discriminating (even for
low-symmetry lattices) relative to similar analyses from 2D data, i.e. from
a only a single image or diffraction pattern. After a basis triplet is
selected, the indices $(hkl)$ of any observed spot $\langle g|$ in a
diffraction pattern follows from $\langle hkl|=\langle g|W$. Similarly,
indices $[uvw]$ of any observed lattice periodicity $|r\rangle $ in an image
follow from $|uvw\rangle =W^{-1}|r\rangle $, once a basis triplet is in
hand. Given the triplet one can similarly calculate goniometer settings to
align the beam with any other crystallographic zone of interest.

\section{The Experimental Setup}

\subsection{Instruments}

The Philips EM430ST TEM used is housed in a triple-bi/story building
designed for low vibration, and provides continuous contrast transfer to $%
\sim 1/(0.19nm)$ at Scherzer defocus. It is equipped with a $\pm 15^{\circ }$
side-entry goniometer specimen stage. A Gatan double tilt holder enables $%
\pm 10^{\circ }$ tilt around an orthogonal tilt axis. The largest
orientation difference which can be achieved using this double tilt holder
in the microscope is therefore $35.6^{\circ }$ \cite{Selb31}.

\subsection{Determining the angle of effective tilt projected onto an image}

In order to establish the spatial relationship between reciprocal lattice
vectors inferred from images taken at different specimen tilts, the
direction of the tilt axis relative to those images must be known. The tilt
axis direction is defined via the right hand rule, as orthogonal to the
relative motion of parts of the specimen as the goniometer reading is
increased. In a single tilt, the axis is perpendicular to the electron beam
and parallel to the micrographs. This is true also of the effective tilt
axis in a double tilt, provided the two specimen orientations are symmetric
about the zero tilt position. We limit our discussion of double tilts to
this case.

We determined the projection of both tilt axes of a Gatan double tilt holder
onto 700K HREM images by examining Kikuchi line shifts during tilt in the
1200mm diffraction pattern of single crystal silicon, and then correcting
for the rotation between that diffraction pattern and the image \cite
{QinThesis}. To be specific, with a micrograph placed in front of the
operator with emulsion side up as in the microscope, with zero azimuth
defined as a vector from left to right, and with counterclockwise defined as
the direction of increasing azimuth, the projection of the $T_{1}$ axis on
1200mm camera-length diffraction patterns in our microscope is along $%
114.0^{\circ }$. The rotation angle between electron diffraction patterns at
the camera length of 1200mm and 700K HREM images is $42.9^{\circ }$.
Therefore the direction of the projection of $T_{1}$ on 700K HREM images is
along $-156.9^{\circ }$. The direction of the projection of the second tilt
axis, $T_{2}$, on 700K HREM images is along $113.1^{\circ }$.

\subsection{A reference coordinate system}

\begin{figure}
\includegraphics{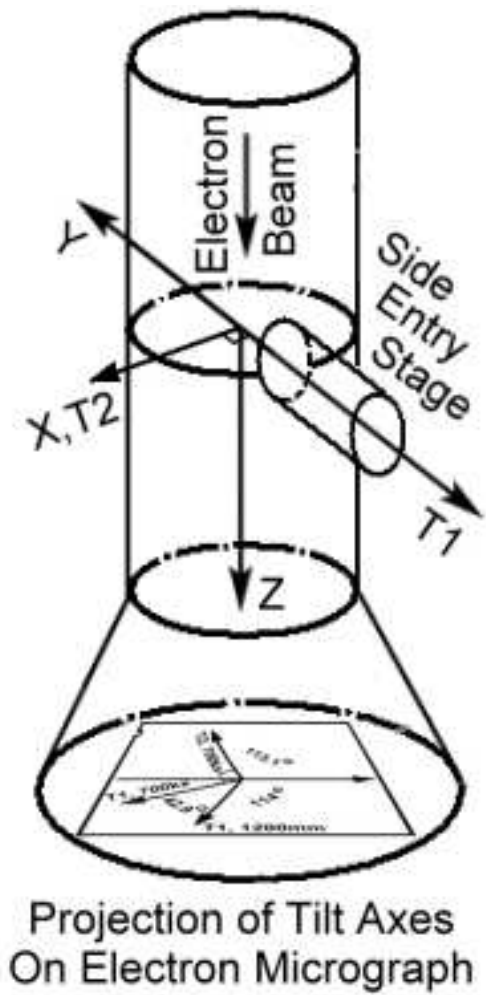}%
\caption{A schematic illustration of the coordinate system set-up for
measuring reciprocal lattice vectors. The coordinate system is fixed to the
microscope. The y and z axes are defined to be along -T$_{1}$ and the
electron beam direction, respectively. The projections of T$_{1}$ and T$_{2}$ 
on 700kx magnification micrographs, as well as of T$_{1}$ on 1200mm diffraction 
patterns, are shown as well.}
\label{Fig4}
\end{figure}

\begin{figure*}
\includegraphics{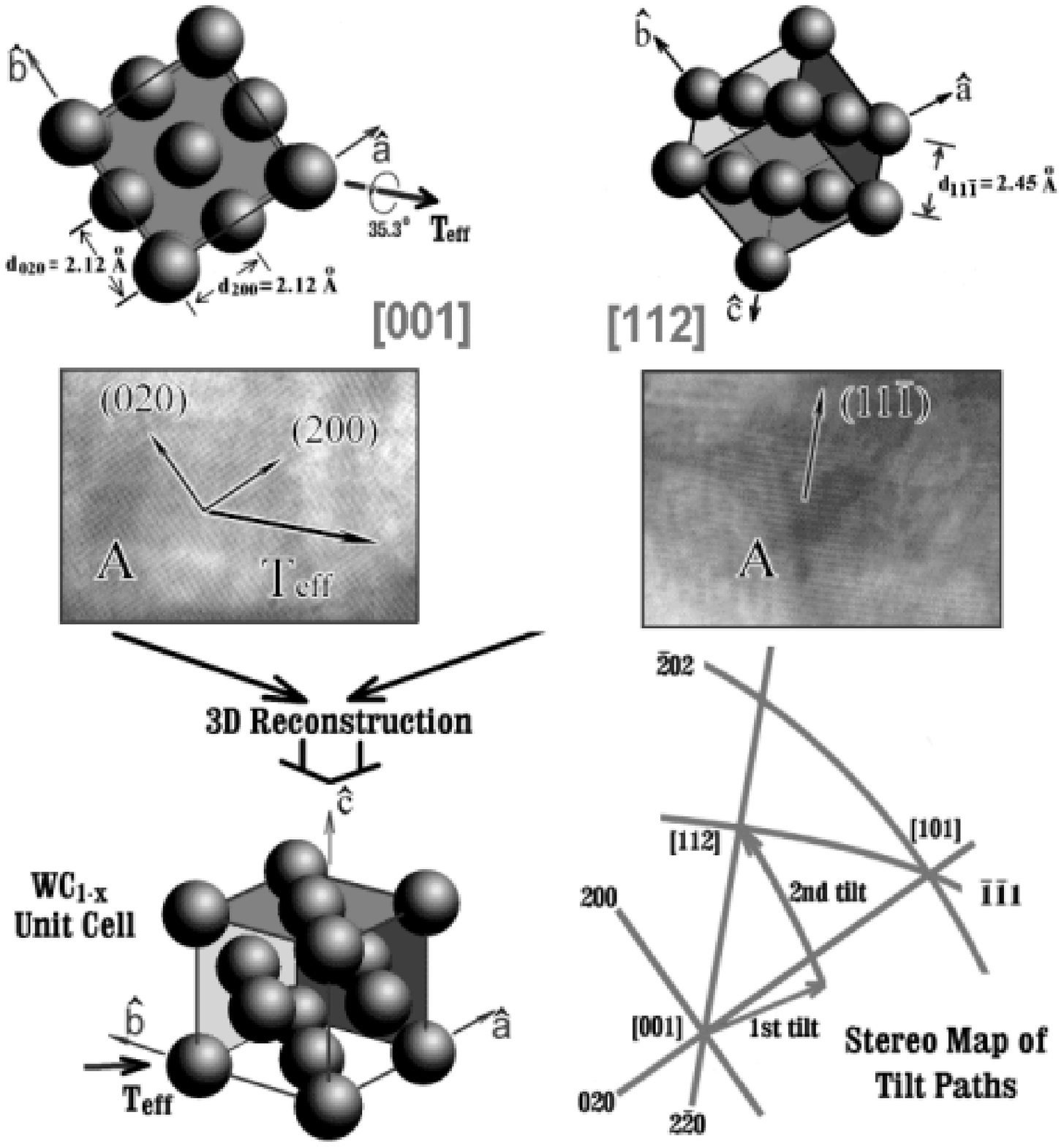}%
\caption{The tilt protocol for inferring the lattice structure of f.c.c. WC$%
_{1-x}$ by viewing a WC$_{1-x}$ crystal from its [001] zone and [112] zone,
along which the two most easily resolved lattice plane sets, the \{200\} and
\{111\} lattice planes, will show lattice fringes in HRTEM images. The two
zones are 35.3$^{\circ}$ apart, within the tilting limit of the microscope
with a Gatan double tilt holder. The crystal is to be viewed along its [001]
zone in the first specimen orientation, then tilted along the ($2\overline{2}%
0$) lattice planes to the [112] zone. The ($2\overline{2}0$) lattice planes
must therefore be perpendicular to the effective tilt axis. The projections
along the [001] and [112] zones together with the effective tilt axis have
been drawn so that their azimuths are consistent with those in the 700K
HRTEM images. The experimental HRTEM images of a WC$_{1-x}$ nano-crystal
acquired using this protocol and their power spectra are shown in the
bottom. The actual tilt sequence is to tilt along -T$_{2}$ by 19.5$^{\circ}$
followed by tilting along -T$_{1}$ by 30.0$^{\circ}$, where T$_{1}$ and T$%
_{2}$ denote the side-entry goniometer tilt axis and the second tilt axis of
a Gatan double tilt holder, respectively.}
\label{Fig1}
\end{figure*}

We then consider a coordinate system fixed to the microscope, for measuring
reciprocal lattice vectors from the power spectra of 700K HREM images. The $%
y $ and $z$ directions are defined to be along $-T_1$ and the electron beam
direction, respectively, as shown in Figure \ref{Fig4}. The projection of
these tilts on the power spectrum of a 700K HREM image is shown in the 2nd
inset of Figure \ref{Fig1}. Azimuths in the remainder of this paper are all
measured in the $xy$ plane of this coordinate system, with the $x$ or $T_2$
direction set to zero. Because the $T_1$ direction is defined in our
coordinate system as the negative $y$-direction, azimuthal angles are
measured on micrographs from a direction $90^{\circ }$ clockwise from the $%
T_1$ direction, when the emulsion side is up.

\subsection{Double tilting}

The specimen was first tilted about $T_2$ to $\vartheta _2=9.7^{\circ }$
while $\vartheta _1$ remained at $0^{\circ }$, made eucentric, then tilted
about $T_1$ to $\vartheta _1=15.0^{\circ }$. The first HREM image was taken
at this specimen orientation of ($\vartheta _1=15^{\circ }$,$\vartheta
_2=9.7^{\circ }$). A similar sequence was applied to take a second HREM
image at the second specimen orientation of ($\vartheta _1=-15^{\circ }$, $%
\vartheta _2=-9.7^{\circ }$). The process can be modeled with a simple
matrix calculation \cite{Liu5,Tamb6}.

Because of the importance of repeatable quantitative tilts, effects of
``mechanical hysterisis'' were minimized by inferring all relative changes
in tilt from goniometer readings taken with a common direction of goniometer
rotation. The rotation sequences were ``initialized'' by first tilting past
the starting line, and then returning to it in the direction of subsequent
motion. Nonetheless both the precision of angle measurement, and our
inability to observe lattice fringes {\em during} rotation, were
shortcomings that microscopes designed to apply these strategies routinely
must address.

\subsection{Specimen preparation}

The tungsten carbide thin film was deposited by PECVD on glass substrates by
introducing a gaseous mixture of tungsten hexacarbonyl and hydrogen into a
RF-induced plasma reactor at a substrate temperature of $330^{\circ}$C \cite
{Qin30}. The specimen was disk-cut, abraded from the glass substrate side and
dimpled by a Gatan Model 601 Disk Cutter, a South Bay Technology Model 900
Grinder and a Gatan Model 656 Precision Dimpler, respectively. The specimen
was then argon ion-milled by a Gatan DuoMill for about 5 hours to
perforation prior to the TEM study, at an incidence angle of $3^{\circ}$.

\section{Experimental Results}

\subsection{Diffraction from a known to check column geometry}

Calibration of these algorithms with geometry in our microscope was first
done with diffraction data from a Si crystal. Diffraction patterns of $%
\langle 100\rangle $ Si along the [$1\overline{1}\overline{6}$] and [$1%
\overline{1}6$] zone axes were obtained by tilting about $T_{1}$ and $T_{2}$%
. The lattice parameters determined are ($a=0.383$nm, $b=0.387$nm, $c=0.386$%
nm, $\alpha =60.0^{\circ }$, $\beta =119.6^{\circ }$, $\gamma =119.1^{\circ
} $). This set of chosen basis defines the rhombohedral primitive cell of
the Si f.c.c lattice. Compared with the literature values of Si ($a=0.384$%
nm, $b=0.384$nm, $c=0.384$nm, $\alpha =60^{\circ }$, $\beta =120^{\circ }$, $%
\gamma =120^{\circ }$), the angular disagreements are less than $1^{\circ }$
and spatial disagreements are less than $1\%$. These uncertainties are
comparable to those obtained by other techniques of submicron crystal
analysis \cite{Liu1,Liu3,Liu5,Tamb6,Phil7}.

\subsection{Nanocrystal images to infer lattice parameters of an unknown}

The micrographs in Figure \ref{Fig1} show a nanocrystal in a film rich in
tungsten carbide, at the orientations of ($\vartheta _{1}=15^{\circ }$, $%
\vartheta _{2}=9.7^{\circ }$) and ($\vartheta _{1}=-15^{\circ }$, $\vartheta
_{2}=-9.7^{\circ }$) respectively.  The coordinates of periodicities in 
micrograph power spectra, as well as in the common reference coordinate system, 
are listed in Table \ref{Table3}, in much the same format as is diffraction 
data for stereo analysis \cite{Phil7}.



\begin{table}
\caption{The g-spacings $[nm^{-1}]$ and azimuths $[^{\circ}]$ of three 
spots measured from the power
spectra of images at two tilts, as well as calculated Cartesian coordinates
$[nm^{-1}]$ of those reciprocal lattice spots in a common coordinate system. This
constitutes a minimal data set for analyzing the lattice in
three-dimensions. }
\begin{tabular}{lccccccc}
Spot $n$ & $g_m$ & $\varphi _m$ & $\vartheta_1$ & $\vartheta_2$ & $g_x$ & $g_y$ & $%
g_z$ \\ 
\hline
1 & 4.73 & 79.2 & -15.0 & -9.7 & 0.861 & 4.53 & -1.01 \\ 
2 & 4.77 & -11.6 & -15.0 & -9.7 & 4.52 & -1.15 & -1.03 \\ 
3 & 4.14 & 32.6 & +15.0 & +9.7 & 3.37 & 2.04 & 1.27 \\ 
\end{tabular}
\label{Table3}
\end{table}

\subsubsection{\bf Matching the lattice:}

\begin{figure}
\includegraphics{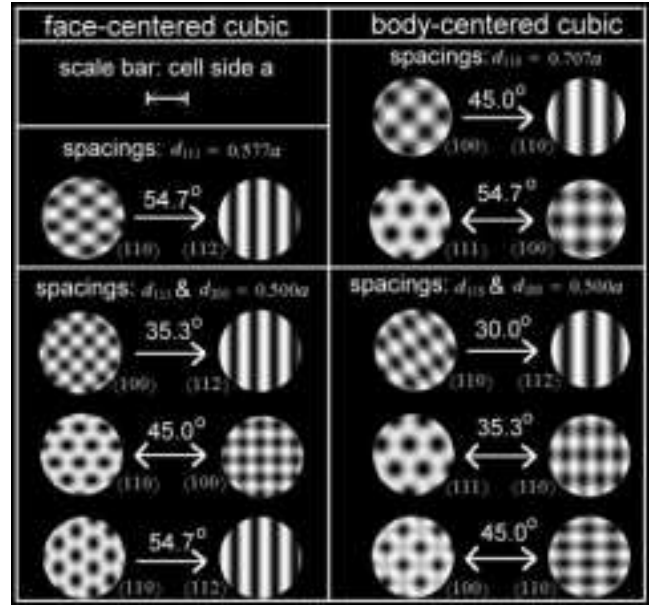}%
\caption{Illustration of all ways (from Figure \ref{Fig0}) to verify the
three dimensional lattice parameters of fcc and bcc crystals from a pair of
lattice images, given an ability to image only lattice spacings down to half
the unit cell side, and a tilt range of less than 60 degrees. The image
pairs are rotated so as to illustrate the {\em direction of tilt} between
the two zones (crystal orientations) involved. The protocols shown with
double arrows provide access to cross-fringes (and hence information on
which direction to tilt) at both ends of the tilt sequence. Also note that
the first and last entries in each column simply provide low and high
resolution views, respectively, of the same experiment.}
\label{BigFig}
\end{figure}

The lattice spacings and inter-spot angles of periodicities in image power
spectra were used to look for consistent indexing alternatives from a set of
36 tungsten carbide and oxide candidate lattices including WC$_{1-x}$. When
an angular tolerance of $2^{\circ }$ and a spatial tolerance of $2\%$ are
imposed, only WC$_{1-x}$ provides a consistent indexing alternative. As
summarized in Table \ref{Table4}, the Miller indices of the three observed
spots then become ($200$), ($020$) and ($11\overline{1}$). The images of Figure \ref{Fig1} thus represent WC$_{1-x}$
[001] and [112] zones, respectively. The azimuth of the reciprocal lattice
vector ($2\overline{2}0$),

\begin{equation}
\varphi _{(2\overline{2}0)}=\frac{\left( \varphi _{(200)}+\left[ 180^{\circ
}+\varphi _{(020)}\right] \right) }{2}=123.8^{\circ }\text{,}
\label{phieqn}
\end{equation}
deviates from the projection of the effective tilt axis by only $0.3^{\circ
} $. Therefore the ($2\overline{2}0$) lattice planes are perpendicular to
the effective tilt axis as per Figure \ref{BigFig}, and the data acquired
are consistent with the expectation for fcc crystals outlined in Figure \ref
{Fig1}. The actual tilting path in the Kikuchi map
of crystal A is shown there as well.

\begin{table}
\caption{The results of a three-dimensional match of measured spacings $[nm]$ and
interspot angles $[^{\circ}]$, with those predicted from the literature 
(denoted with a
caret) for the face-centered cubic crystal WC$_{1-x}$. This phase is the
only one from a list of 36 tungsten carbides and tungsten oxides whose
predicted spacings and interspot angles agreed the measurements within a
tolerance of $1.5\%$ and $1.5^{\circ }$, respectively.}
\begin{tabular}{lccccccc}
Spot $n$ & $d_n$ & $\varphi _{ij\neq n}$ & $(hkl)$ & $%
\widehat{d_{hkl}}$ & $\frac{\delta d}d[\%]$ & $\widehat{\varphi _{ij\neq
n}}$ & $\delta \varphi$ \\ 
\hline
1 & 0.212 & 54.2 & ($200$) & 0.212 & 0.5 & 54.7 & 0.5 \\ 
2 & 0.209 & 56.2 & ($020$) & 0.212 & 1.4 & 54.7 & 1.5 \\ 
3 & 0.242 & 90.8 & ($11\overline{1}$) & 0.245 & 1.2 & 90.0 & 0.8 \\ 
\end{tabular}
\label{Table4}
\end{table}

From the indexing suggested by this match, the $x$, $y$, and $z$ coordinates
of the reciprocal lattice basis vectors $a^{\ast }$, $b^{\ast }$, and $%
c^{\ast }$ may be inferred. This is shown in Table \ref{Table6}, along with
the resulting lattice parameters and a comparison with literature values.
The resulting errors in $a$ and $b$ are less than $1.3\%$, while the error
in $c$ (which is orthogonal to the plane of the first image) is larger
(around $2.3\%$). Both because of tilt uncertainties and reciprocal lattice
broadening in the beam direction, uncertainties orthogonal to the plane of
the specimen are expected to be larger than in-plane errors \cite{Phil27}.

\begin{table}
\caption{Primitive (top) and face-centered (bottom) reciprocal lattice basis
triplets (spacings in nm, angles in degrees) inferred from our two HREM images of crystal A, along with a
comparison of the lattice parameters for each cell which follow therefrom.
The indexing of the observed reciprocal lattice vectors in these two cases
was inferred directly for the primitive cell, by minimizing cell volume, and
by matching to ``textbook'' parameters for WC$_{1-x}$ in the case of the
face-centered cell. Except for the choice of basis triplet, the two measured
cells refer to exactly the same inferred lattice.}
\begin{tabular}{lcccccc}
{\em measured} & 0.298 & 0.299 & 0.296 & 120.0 & 58.7 & 119.8 \\ 
{\em ``book''} & 0.300 & 0.300 & 0.300 & 120 & 60 & 120 \\ 
\hline
{\em params} & $a$ & $b$ & $c$ & $\alpha $ & $\beta $ & $\gamma $
\\ 
\hline
{\em ``book''} & 0.425 & 0.425 & 0.425 & 90 & 90 & 90 \\ 
{\em measured} & 0.424 & 0.419 & 0.415 & 88.5 & 90.8 & 89.3 \\ 
\end{tabular}
\label{Table6}
\end{table}

\subsubsection{\bf Building a triplet from scratch:}

By generating linear integral combinations of the measured periodicities in
reciprocal space (i.e. vector triplets of the form $%
n_{1}g_{1}+n_{2}g_{2}+n_{3}g_{3}$ where the $n_{i}$ are integers) until a
minimal volume unit cell is obtained (there will be more than one way to
achieve the minimum), a primitive triplet for the measured lattice can be
inferred quite independent of any knowledge of candidate lattices. The
primitive cell parameters determined are also listed in Table \ref{Table6}.
With respect to literature values, these show spatial disagreements of less
than $1.6\%$, and angular disagreements of less than $1.5^{\circ }$.
Although inference of the ``conventional cell'' from the primitive cell
alone is possible, the process has not been attempted here because of
complications attendant to measurement error.

\subsubsection{\bf Phase Identification}

Determining a reciprocal lattice triplet, and inferring lattice parameters
therefrom, are of course not equivalent to confirming the existence of a
particular phase. In order to draw a more robust conclusion about the makeup
of crystal A, we extended our analysis of the structure to other lattices
capable of indexing the observed spots, albeit with larger errors in spacing
and interspot angle. When the spatial and angular tolerances of our
candidate match analysis are increased to $3^{\circ}$ and $3\%$, there are
many tungsten oxide and carbide candidates in addition to WC$_{1-x}$ which
show consistent lattice spacings and inter-planar angles \cite{Qin30}.

In order to eliminate these candidates, it was necessary to confirm, using
power spectra of amorphous regions in each image, that the spatial
frequencies in Figure \ref{Fig1}images were continuously
transferred within the first passband \cite{Spen36,Will37}. By then assuming
that projected reciprocal lattice frequencies make their way into the exit
surface wavefield (at least at the thin edges of the particle), all the
candidates except WC$_{1-x}$ are eliminated. Specifically, it was found that
for each of the candidates except WC$_{1-x}$, along one or more of the
suggested zone axes at least one reciprocal lattice vector shorter than the
experimental one(s) is missing in a power spectrum \cite{Qin30}. An example
of this is the match with hexagonal WC$_x$ ($a=1.058$nm, $c=1.335$nm). In
this case the Miller indices suggested for spot 3 ($\overline{4}2\overline{2}
$) were inconsistent with the fact that the ($\overline{2}1\overline{1}$) is
absent from the power spectrum of the image on the right side of Figure \ref{Fig1}.

Our conclusion that ``this crystal is WC$_{1-x}$'' (and as we see later most
of the other crystals in the specimen) is consistent with knowledge of the
formation conditions, as well as with X-ray powder and EDS analysis of the
same film.

\subsection{The effective tilt direction, and recurring fringes}

In addition to serving as a guide for correctly choosing the azimuth of the
crystal before tilting between desired zones, knowledge of the tilt axis
direction plays another role: that of highlighting lattice fringes present
in both specimen orientations, but caused by one and the same set of lattice
planes.

In single tilt experiments, the tilt axis is simply $T_{1}$. This is always
perpendicular to the electron beam and hence parallel to the micrographs.
Any reciprocal lattice vector parallel or antiparallel to $T_{1}$ remains in
Bragg condition throughout the whole tilting process, regardless
of the amount of tilt $\vartheta _{1}$. If the spacing is large enough to be
recorded in the images, the same lattice fringes are seen perpendicular to
the projection of $T_{1}$ in any HREM image as well.

For double tilt experiments, it is convenient to introduce the concept of an 
{\em effective tilt axis}. The effective tilt axis is analogous to the tilt
axis in a single tilt experiment, in that the double tilt can be
characterized by a single tilt around the effective tilt axis of angular
size equal to that in the double tilt. This effective axis is perpendicular
to the electron beam and hence parallel to the micrographs only if the two
specimen orientations are symmetric about the untilted position.

Considering only double tilts falling into this category, let ($\vartheta
_{1}$, $\vartheta _{2}$) and ($-\vartheta _{1}$, $-\vartheta _{2}$) denote
the specimen orientations before and after. The effective tilt axis
direction has an azimuth (with respect to our reference $x$-direction) of

\begin{equation}
\varphi _{eff}=\tan ^{-1}\left[ -\frac{\sin (\vartheta _{1})}{\tan
(\vartheta _{2})}\right] \text{.}  \label{eTilt}
\end{equation}
A proof of equation \ref{eTilt} is given in Appendix A. There
exists a 180$^{\circ }$ ambiguity in the direction of the effective tilt
axis using equation \ref{eTilt}. This ambiguity can be resolved through the
knowledge of the actual tilting sequence. In our experiment $\vartheta
_{1}=15^{\circ }$, $\vartheta _{2}=9.7^{\circ }$, $\varphi
_{eff}=123.5^{\circ }$. This is the effective tilt axis direction mentioned
in previous sections.

Lattice planes perpendicular to the effective tilt axis, in the double tilt
case, diffract and are visible at initial and final, but not intermediate,
specimen orientations. This result inspired further experimental work on,
and modeling of, fringe visibility loss during tilt \cite{QinThesis}. One
result of this exercise was a prediction that $0.213$nm fringes deviating by
as much as $4^{\circ }$ from the effective tilt direction in a 10nm $%
WC_{1-x} $ specimen will remain visible after a $35.6^{\circ }$ tilt. This
was confirmed by experiment on these specimens \cite{QinThesis}. The result
in turn serves to constrain the probability and error analyses below.

\subsection{Tilt limitations and chances for success}

This section addresses the chances for successful 3D cell determination from
images, depending on properties of both microscope and specimen. Such
matters are considered in more detail elsewhere \cite{QinThesis}.

In a microscope with a single-axis tilt of at least $\pm 35.3^{\circ }$ and
a stage capable also of $180^{\circ }$ rotation, any cubic crystal with an [$%
001$] zone in the beam direction at zero tilt can be re-aligned by azimuthal
rotation until its ($2\overline{2}0$) reciprocal lattice vector is parallel
to T$_{1}$. With an untilted tilt-rotate stage, this would allow a crystal's
[$001$] zone to remain aligned with the beam throughout the rotation.
Subsequent tilting by $35.3^{\circ }$ will lead to the [$112$] zone, and the
lattice structure in three dimensions confirmed ala Figure \ref{Fig1}.

Under these conditions, any cubic crystal showing [$001$] zone cross fringes
can be tilted so as to reveal a third spacing. Hence the probability of
success with any given crystal is that of finding a randomly-oriented
crystal oriented with [$001$]-zone fringes visible. Fortunately for this
method, the spreading of reciprocal lattice points due to finite crystal
thickness $t$ allows one to visualize fringes within a half angle $\Theta
_{t}$ of order $1/t$ surrounding the exact Bragg condition. Otherwise,
cross-fringes would be rare indeed!

The solid angle subtended by this visibility range for lattice planes
intersecting along the [$001$]-zone allows us to calculate the probability
that a randomly-oriented crystal will show cross-fringes of specified type.
For example if we approximate the cross-fringe region with a conical bundle
of directions about each zone, then for the special case of spherical
particles the fraction of crystals showing the fringes of zone $x$ is:

\[
p_{x}=n(1-\cos [\Theta _{t}])\text{, where }\Theta _{t}=\arcsin [\frac{%
g_{t}^{2}-g_{x}^{2}+2g_{\lambda }g_{t}}{\sqrt{2}g_{x}g_{\lambda }}]\text{,} 
\]
$n$ is multiplicity of zone $x$ (e.g. $n=3$ for cubic $x=$[$001$]), $%
g_{x}=1/d$, $g_{\lambda }=1/\lambda $, and $g_{t}=f/t$, where $t$ is the
thickness of the crystal in the direction of the beam and $f$ is a parameter
of order one that empirically accounts for signal-to-noise in the method
used to ``visualize'' fringes. For example, we expect $f$ to decrease if an
amorphous film is superimposed on the crystals being imaged. The half-angle $%
\Theta _{t}$ is a pivotal quantity in both the probability and accuracy of
fringe measurement.

\begin{figure*}
\includegraphics{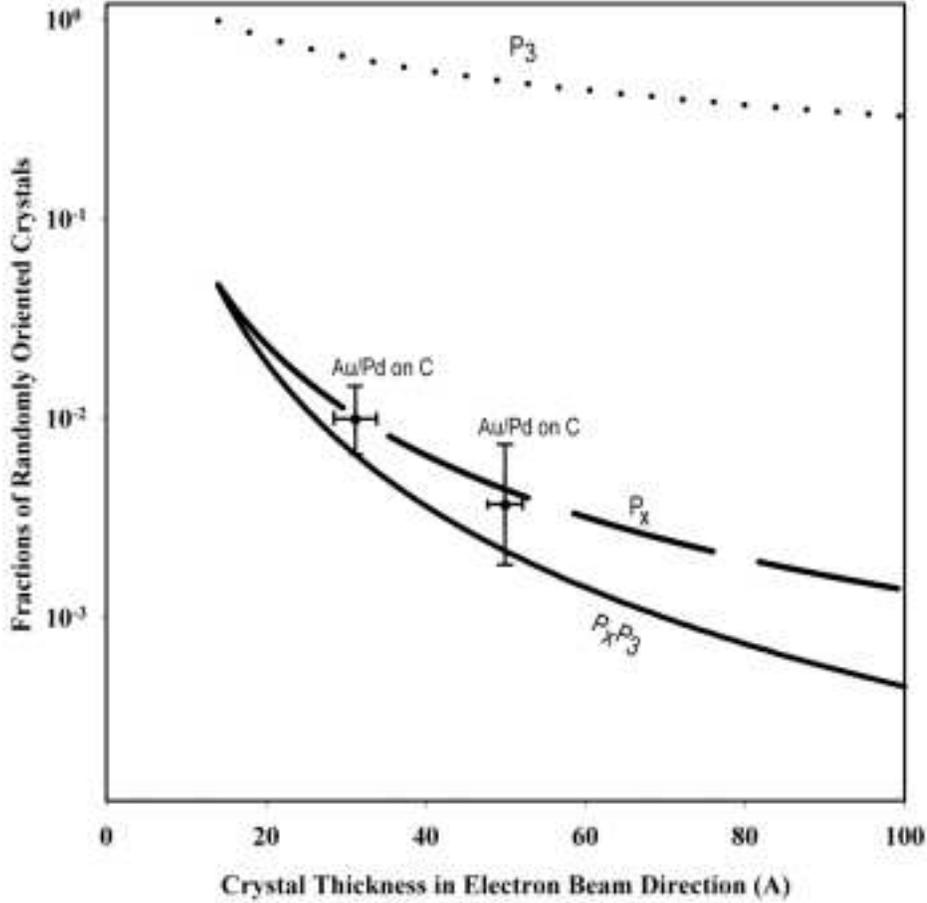}%
\caption{A plot of the fraction of randomly oriented grains of WC$_{1-x}$
showing [001]-zone cross fringes ($p_x$), and the fraction of such
cross-fringe grains oriented so that a random tilt of 35.3 $^{\circ}$ will
allow imaging of a 111 periodicity ($p_3$), as a function of specimen
thickness in the direction of the electron beam. \protect\cite{PaperC}}
\label{prob}
\end{figure*}

A plot of the probability for seeing [$001$] cross-fringes of spacing $%
d=0.202$nm, as a function of specimen thickness $t$ for both spherical and
laterally-infinite (rel-rod) particles, is shown in Fig. \ref{prob}. Here
we've used fit parameter $f=0.79$ based on data points (also plotted) that
were obtained experimentally for particles of varying size from HREM images
of Au/Pd evaporated onto a carbon film. As you can see, the probability of
encountering cross-fringes improves greatly as crystallite size decreases
toward a nanometer. Of course, as discussed in the next section, this
``reciprocal lattice broadening'' is accompanied by a decrease in the
precision of measurements for individual lattices.

Due to the tilt limits of the specimen holder in our microscope, the first
HREM image along the [$001$] zone of a WC$_{1-x}$ nano-crystal had to be
taken at a nonzero $\vartheta _{1}$ orientation. Azimuthal symmetry is thus
broken. Our solution was to find a [$001$] nano-crystal whose ($2\overline{2}%
0$) reciprocal lattice vector was by chance parallel to the effective tilt
axis, then tilting to the 2nd orientation. Thus nano-crystal $A$ was
identified (by coincidence) to have an appropriate azimuth during real time
study of the ($200$) and ($020$) lattice fringes. Tilting by $35.3^{\circ }$
was done thereafter.

For the probability of success in our case, we must multiply $p_{x}$ by the
probability of viewing ($111$) fringes {\em after} tilting a [$001$] crystal
with random aziumth by $35.26^{\circ }$. This probability of finding a 3rd
spacing takes the form $p_{3}=m\delta /\pi $, where $m$ is the multiplicity
of target spacings (e.g. $m=4$ for a four-fold symmetric [$001$] starting
zone), and the ``azimuthal tolerance half-angle'' $\delta $ (again in the
spherical particle case) obeys the implicit relation:

\[
\theta _o=\arctan [\frac{\tan \theta _o}{\cos \delta }]+\arctan [\frac{\cos
\gamma}{\sin ^2\gamma -(\cos \theta _o\sin \delta )^2}]\text{,} 
\]
where $\theta _o$ is the required tilt (in our case $35.26^{\circ }$) and

\[
\gamma =\arccos [\frac{g_{d}^{2}-g_{l}^{2}-2g_{\lambda }g_{l}}{%
2g_{d}g_{\lambda }}]\text{.} 
\]
The probability $p_{3}$ is also plotted as a function of specimen thickness
in Fig. \ref{prob}, along with the product $p_{x}p_{3}$.

These models predict a probability of success with the strategy adopted in
our experiment, for the ``large'' $10$nm WC$_{1-x}$ crystals in our
specimen, of $p_{x}p_{3}=7.3\times 10^{-4}\times 0.371=2.4\times 10^{-4}$.
Hence only one in every $1500$ crystals will show [$001$] cross fringes, and
one in every $4000$ will be suitably oriented for 3D lattice parameter
determination. This is consistent with our experience: The image of crystal
A was recorded in one negative out of 22, each of which provided an
unobstructed view of approximately $100$ crystals.

As mentioned above, using a microscope capable of side-entry goniometer
tilting by $\pm 35.3^{\circ }$ with a tilt-rotate stage, the 3D parameters
of all cubic crystals, when untilted showing [$001$] zone cross-fringes,
could have been determined. According to Fig. 14a, the fraction of particles 
$2$ nm in thickness that are oriented suitably for such analysis approaches $%
1$ in $100$. Moreover, with a goniometer capable of tilting by $\pm
45^{\circ }$ plus computer support for automated tilt/rotation from any
starting point, {\em each unobstructed nano-crystal} in the specimen could
have been subjected to this same analysis after a trial-and-error search for
accessible [$001$] zones. Thus a significant fraction of crystals in a
specimen become accessible to these techniques, with either a sufficient
range of precise computer-supported tilts, or if the crystals are
sufficiently thin.

\begin{figure*}
\includegraphics{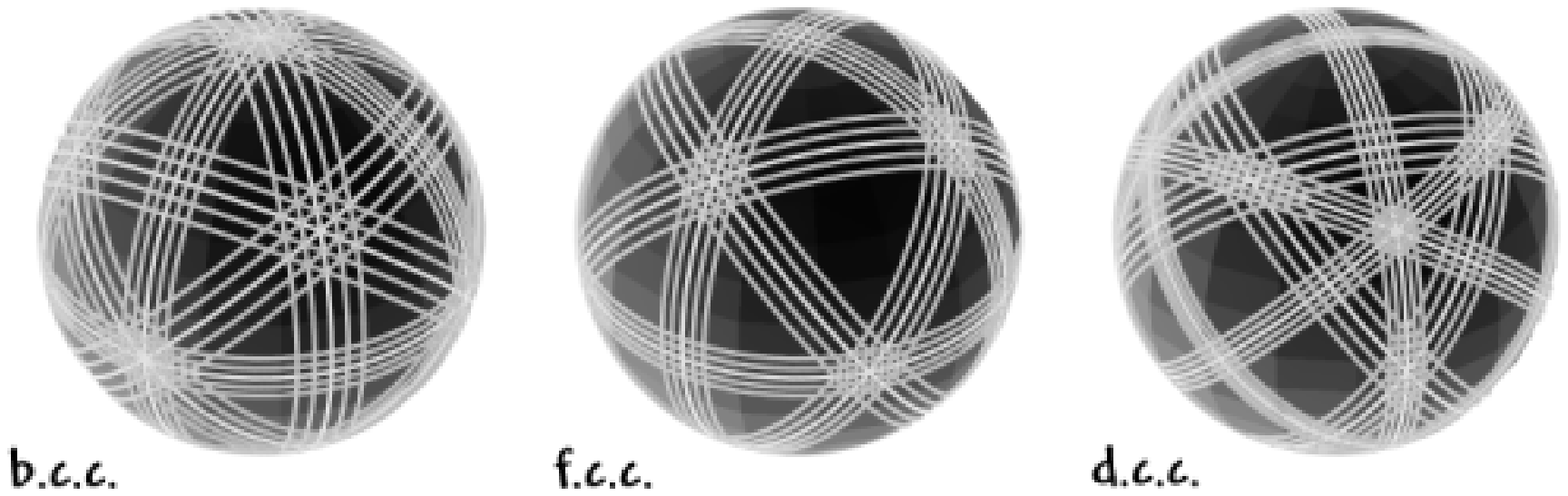}%
\caption{Visibility maps for the two largest fringes visible from body-centered, 
face-centered, and diamond face-centered lattices.  Sphere diameter (relative to band width and fringe spacing) is proportional to specimen thickness, here chosen to be about 5 times the cubic cell side a.  Note the dominance of crossed $(110)$ fringes at the three-fold $\langle 111\rangle$ zone in the body-centered case, the dominant crossed 
$(111)$ fringes at the two-fold $\langle 110\rangle$ zone in the face-centered cases, and the wider disparity 
between largest and next-to-largest spacings when the diamond glide is added to the lattice.}
\label{FigMaps}
\end{figure*}

  Subsequent work \cite{Maps} has shown that information on tilt protocols and 
fringe visibility for crystals of a given thickness can be elegantly 
sumarized with spherical maps like those in Figure \ref{FigMaps}.  
These are a direct-space analog to 
reciprocal-space Kikuchi maps, in which band thickness is proportional 
to d (rather than 1/d).  If the sphere used has a diameter equal to 
specimen thickness, then the first order effect of changing thickness 
simply increases the separation between zones while holding the width 
of the bands fixed.

\section{Pitfalls and uncertainties}

\subsection{Cautions involving specimens and contrast transfer}

In this section, we discuss effects warranting caution. In the next section,
models of lattice parameter uncertainty are discussed.

High electron beam intensities can cause lattice rearrangement in
sufficiently small nanocrystals, as well as changes in the orientation of a
thin film (e.g. due to differential expansion). Sequential images of the
same region at fixed tilt might allow one to check for such specimen
alterations.

Loss of periodicities in the recorded image, due to damping and spherical
aberration zeros, were discussed in the section above on experimental
design. Nonetheless, careful observations of more than one crystal, and
image simulation as well, may be useful adjuncts whenever this technique is
applied. We illustrate this below, with a ``two-dimensional'' experiment
done to assess the size of errors due to finite crystal size and random
orientation. The result is of help in the section on modeling uncertainties
that follows.

\begin{figure*}
\includegraphics{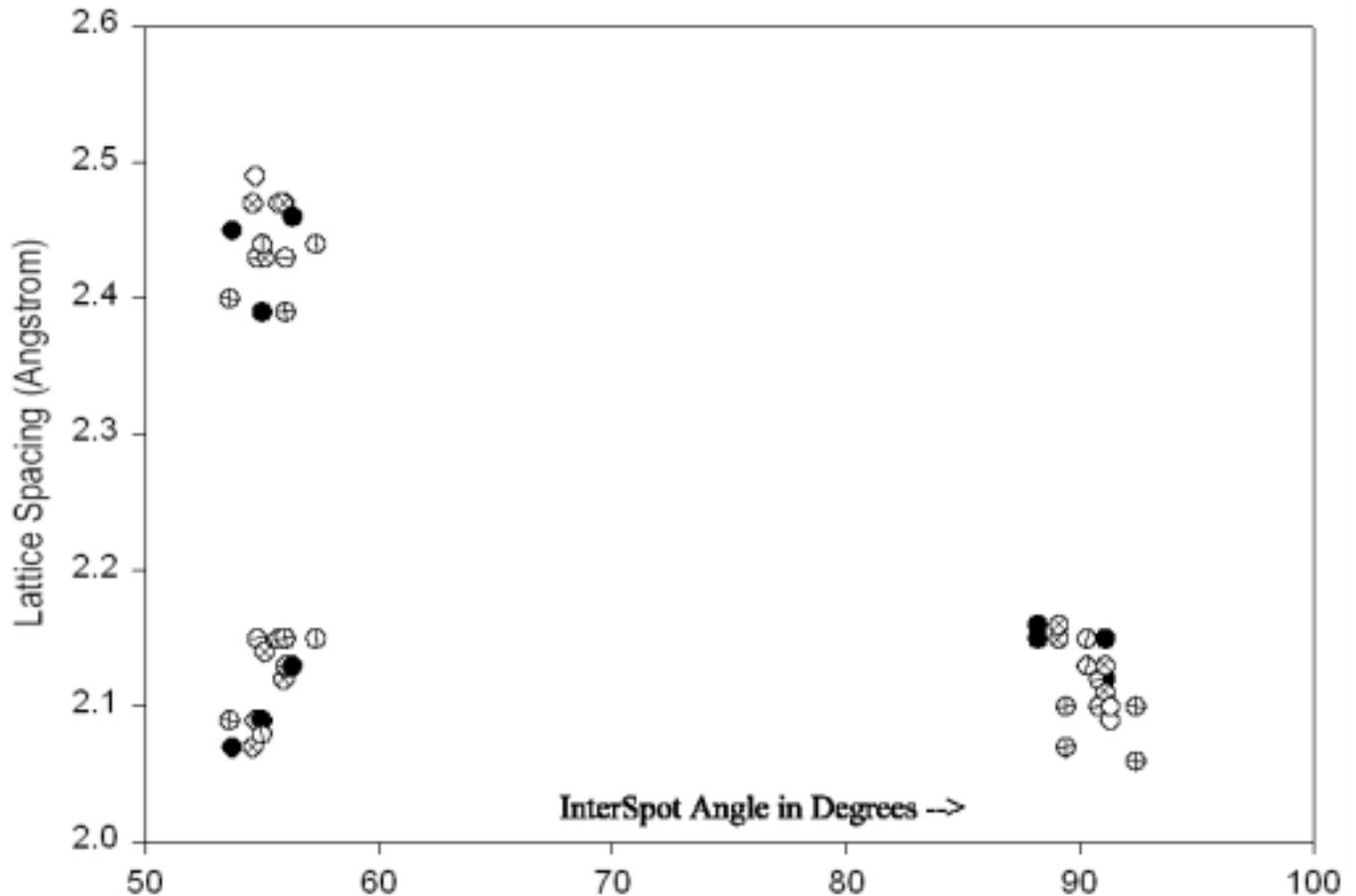}%
\caption{The spacings and interplanar angles measured from the cross lattice
fringes of 23 nano-crystals free of overlap with other crystals in the HRTEM
images. Insets (b) and (c) are magnified plots of the two regions in (a)
where all the data are concentrated. The specific combinations of lattice
spacings and interplanar angles corresponds to the [001] and [011] zone
images of WC$_{1-x}$ and hence indicate WC$_{1-x}$ being the only present
phase in the thin film. The measured lattice spacings have a standard
deviation from their mean of less than 1.5\%, and interplanar angles have a
standard deviation of 1.2$^{\circ}$.}
\label{Fig12}
\end{figure*}

A recent paper on HREM image simulations \cite{Malm39} indicated that
deviations in orientation of a 2.8nm spherical palladium nano-crystal from
the zone axes may result in fringes unrelated to the structure of the
particle. Variability in measured lattice spacings was also reported to be
as high as several percent, with the highest reaching 10\%. To compare such
results with our experimental data, 23 single crystals free of overlap with
other crystals, and each showing cross-fringes, were examined. The projected
sizes of these crystals range from 3.7nm$\times $3.8nm to 10.8nm$\times $%
7.8nm. The spacings and angles between fringes are plotted in Fig. \ref
{Fig12}.

Observed cross-fringes in the HREM images fall into two categories,
according to their spacings and angles. The first category is characterized
by a $90^{\circ}$ inter-planar angle between two $2.12\AA $\ lattice
spacings. The second one by two inter-planar angles of $55^{\circ }$, $%
70^{\circ }$ and three lattice spacings of $2.12\AA $, $2.12\AA $, $2.44\AA $%
. Only the spacings of $2.44\AA $\ and $2.12\AA $\ and the corresponding
angle of $55^{\circ }$ have been shown in Figure \ref{Fig12}. Two
conclusions can be drawn.

First, since the two categories of cross-fringes match those along the [$001$%
] and [$011$] zone axes of WC$_{1-x}$, the only two zones which show
cross-lattice fringes in our HREM images, the thin film consists mainly of WC%
$_{1-x}$. X-ray powder diffraction work on the film supports this
conclusion \cite{Qin30}.

Secondly, for nano-crystals free of overlap with other crystals, the
observed lattice spacings and interplanar angles in HREM images have a
standard deviation from the mean of less than $1.5\%$, and a standard
deviation of less than $1.3^{\circ}$, respectively. We have not observed any
seriously shortened or bent fringes. Nonetheless we recommend that such
fringe abundance analyses go hand in hand with stereo lattice studies of
nano-crystal specimens, and that comparative image simulation studies be
done where possible as well.

\subsection{Uncertainty forecasts}

Earlier estimates \cite{Phil27}, as well as the typical size of diffraction
broadening in the TEM, suggest that lattice parameter spacing errors may in
favorable circumstances be on the order $1\%$, and angle errors on the order
of $1^{\circ }$. Experiment, and a more detailed look at the theory \cite
{QinThesis}, now support this impression. We will focus the discussion here
on equant or spherical nanocrystals. The results should be correct within $%
10\%$ for other (e.g. thin-foil) geometries of the same thickness.

The three sources contributing to the lattice spacing measurement
uncertainty in images are: expansion of the reciprocal lattice spot in the
image plane, uncertainty in the camera constant, and expansion of the
reciprocal lattice spot along the electron beam direction, in order of
decreasing relative effect \cite{QinThesis}. The uncertainty in our camera
constant is measured to be about $0.5\%$. Uncertainties from the first and
third sources above are on the order of $1\%$ and $0.01\%$, respectively,
for a typical lattice spacing of 0.2 nm. The in-plane/out-of-plane error
ratio is on the order of 10.

Sources contributing to the measurement uncertainty of lattice parameters
along the electron beam direction, when the specimen is un-tilted, include
uncertainty in goniometer tilt as well as sources analogous to those above.
Observation of reciprocal lattice vectors further out of the specimen plane
(i.e. of fringes at high tilt) reduces the measurement uncertainty of
``out-of-plane'' parameters. \ The measurement uncertainty of inter-planar
angles in images is due to lateral uncertainty in their associated
reciprocal lattice spots in the image plane.

Using a mathematical model of these errors \cite{QinThesis}, we predict
spacing uncertainties in a 10 nm nanocrystal, tilted by $\pm 18^{\circ }$,
of $2.1\%$ for an imaged spacing and of $8.6\%$ for a lattice parameter
perpendicular to the plane of the untilted specimen. This large
``out-of-plane'' uncertainty is a result of the small tilt range available
with our high resolution pole piece. The estimated interplanar angle
uncertainty is about $2.3^{\circ }$. These predicted uncertainties \cite
{QinThesis} are between 2 and 3 times the errors observed here, and hence of
the right order of magnitude.

The model suggests that lattice parameter uncertainties will decrease as
camera constant and tilt uncertainties decrease, and will also decrease as
the tilt range used for the measurement increases. It suggests that the
lattice parameter errors will increase as crystal thickness goes down. The
ease of locating spacings, however, goes up as crystal thickness decreases.
Hence the best candidates for application of the protocols here may be
crystals in the 1 to 20 nm range. Improved tilt accuracy (hopefully with
computer guidance), and low-vibration tilting so that fringes may be
detected as orientation changes, would make these strategies more accurate
and widely applicable as well.

\section{Conclusions}

When considered from the perspective of direct space imaging, crystals offer
a discrete set of opportunities for measuring their lattice parameters in
three dimensions. Ennumerating those opportunities for candidate lattices,
or lattice classes, opens doors to the direct experimental determination of
nano-crystal lattice parameters in 3D. A method for doing this, and lists of
those opportunities for the special case of cubic crystals, are presented
here.

We apply this insight to inferring the 3D lattice of a single crystal from
electron phase or Z-contrast images taken at two different orientations. For
nano-crystals in particular, a double-axis tilt range of $\pm 18^{\circ }$
degrees allows one to get such data from all correctly-oriented cubic
crystals with appropriate spacings resolvable in a pair of images taken from
directions separated by $35.3^{\circ }$. In the experimental example
presented, we find less than $1.5\%$ spatial and $1.6^{\circ }$ angular
disagreements between the inferred primitive cell lattice parameters of a 10
nm WC$_{1-x}$ nano-crystal, and literature values.

We further present data on the variability of lattice fringe spacings
measured from images of such randomly-oriented 10nm WC$_{1-x}$ crystals in
electron phase contrast images. The results suggest that measurement
accuracies of $2\%$ in spacing and $2^{\circ }$ in angle may be attainable
routinely from particles in this size range. Smaller size crystals may be
easier to obtain data from, but show larger uncertainties, while larger or
non-randomly oriented crystals (especially if guesses as to their structure
are unavailable) may be more challenging to characterize in three dimensions.

Precise knowledge of the tilt axes, as projected on the plane of a
micrograph, is crucial to implementation. This information, if coupled with
on-line guidance on how to tilt from an arbitrary two-axis goniometer
orientation in any desired direction with respect to the plane of an image
or diffraction pattern, could make this strategy and related diffraction
strategies \cite{Phil7} for lattice parameter measurement more routine.
Future microscopists might then be able to interface to individual 
nanocrystals much as the nano-geologist in the first paragraph of this 
paper examined her ``hand specimen''.

Lastly, diffraction can also be used in this ``stereo analysis'' mode (with crystals 
large enough to provide diffraction patterns), although the easier accessibility of
high spatial frequencies via diffraction often makes the large tilts used
here unnecessary.  They can also be put to use in darkfield imaging
applications, by forming images of the specimen using ``beams'' diffracted
by the periodicities which serve as diagnostic of a given lattice (for
example those associated with each of the protocols of Figure \ref{Fig0}).

Images so taken of nano-crystalline specimens, for example at two tilts with
three different darkfield conditions, would be expected to show correlations
among that subset of the crystals correctly oriented for diffraction with
all three reflections. Although this strategy may never allow precise
lattice parameter determinations given limits on objective aperature angular
size, it may be a very efficient way to search for crystals
correctly-oriented and of correct type for one of the imaging protocols
described here. Moreover, because such lattice-correlations in three
dimensions contain information beyond the pair-correlation function, they
may be able to support the new technique of fluctuation microscopy \cite
{Treacy1,Treacy2,Gibson1} in the study of paracrystalline specimens like
evaporated silicon and germanium \cite{Gibson2,Gibson3} whose order-range is
too small for detection by other techniques.

\appendix

\section{Effective tilt axis azimuth}

Let ($\vartheta _{1}$, $\vartheta _{2}$) and ($\vartheta _{1}^{\prime }$, $%
\vartheta _{2}^{\prime }$) denote orthogonal tilt values for two specimen
orientations, and $\phi _{eff}$ the azimuth of the effective tilt axis
between these orientations. Any reciprocal lattice vector with untilted
Cartesian coordinates $|g\rangle $, and with identical micrograph
coordinates $|g_{m}\rangle $ at the two tilted orientations, will (following
equation \ref{g_eqn}) obey:

\begin{equation}
A(\vartheta _1,\vartheta _2)|g_m\rangle =|g\rangle =A(\vartheta _1^{\prime
},\vartheta _2^{\prime })|g_m\rangle \text{,}  \label{a_eqn2}
\end{equation}
where

\begin{equation}
|g_m\rangle =\left( 
\begin{array}{c}
g\cos \left( \varphi _{eff}\right) \\ 
g\sin \left( \varphi _{eff}\right) \\ 
0
\end{array}
\right) \text{.}  \label{qmx2}
\end{equation}

Expanding, this gives three equations which, combined with 
equation \ref{qmx2} $g_{mx}=g_{my}\tan (\varphi _{eff})$, can be 
solved for the three unknowns $\vartheta _1^{\prime }$%
, $\vartheta _2^{\prime }$, $\varphi _{eff}$, to get:

\begin{equation}
\vartheta _1^{\prime }=-\vartheta _1\text{, }\vartheta _2^{\prime
}=-\vartheta _2\text{, and}  \label{app_eqn6}
\end{equation}

\begin{equation}
\varphi _{eff}=\tan ^{-1}\left[ -\frac{\sin (\vartheta _1)}{\tan (\vartheta
_2)}\right] \text{.}  \label{app_eqn7}
\end{equation}
This provides an equation for the azimuth of the effective tilt, and
confirms that symmetry about the zero tilt position is a necessary and
sufficient condition for the reciprocal lattice vector $|g\rangle $, and
it's associated lattice fringe, to show a common direction in micrographs at
both tilts.

\begin{acknowledgments}
Thanks to W. Shi, J. Li and W. James at University of Missouri - Rolla for 
the tungsten carbide specimens, and to MEMC and Monsanto for regional facility support.
\end{acknowledgments}

\bibliography{temrefs1}

\begin{thebibliography}{25}
\expandafter\ifx\csname natexlab\endcsname\relax\def\natexlab#1{#1}\fi
\expandafter\ifx\csname bibnamefont\endcsname\relax
  \def\bibnamefont#1{#1}\fi
\expandafter\ifx\csname bibfnamefont\endcsname\relax
  \def\bibfnamefont#1{#1}\fi
\expandafter\ifx\csname citenamefont\endcsname\relax
  \def\citenamefont#1{#1}\fi
\expandafter\ifx\csname url\endcsname\relax
  \def\url#1{\texttt{#1}}\fi
\expandafter\ifx\csname urlprefix\endcsname\relax\def\urlprefix{URL }\fi
\providecommand{\bibinfo}[2]{#2}
\providecommand{\eprint}[2][]{\url{#2}}

\bibitem[{\citenamefont{Wyckoff}(1982)}]{Wyckoff}
\bibinfo{author}{\bibfnamefont{R.~W.~G.} \bibnamefont{Wyckoff}},
  \emph{\bibinfo{title}{Crystal Structures}} (\bibinfo{publisher}{Krieger
  Publishing Company}, \bibinfo{address}{Malabar, Florida},
  \bibinfo{year}{1982}), pp. \bibinfo{pages}{7--16}.

\bibitem[{\citenamefont{Qin}(2000)}]{QinThesis}
\bibinfo{author}{\bibfnamefont{W.}~\bibnamefont{Qin}}, Ph.D. thesis,
  \bibinfo{school}{University of Missouri - St. Louis/Rolla}
  (\bibinfo{year}{2000}).

\bibitem[{\citenamefont{Krainer and Robitsch}(1967)}]{Krai28}
\bibinfo{author}{\bibfnamefont{V.~E.} \bibnamefont{Krainer}} \bibnamefont{and}
  \bibinfo{author}{\bibfnamefont{J.}~\bibnamefont{Robitsch}},
  \bibinfo{journal}{Planseeberichte Fuer Pulvermetallurgie}
  \textbf{\bibinfo{volume}{15}}, \bibinfo{pages}{46} (\bibinfo{year}{1967}).

\bibitem[{JCP(1988)}]{JCPD29}
\emph{\bibinfo{title}{PDF-2}}, \bibinfo{type}{Tech. Rep.}
  \bibinfo{number}{25-1316}, \bibinfo{institution}{JCPDS and International
  Centre for Diffraction Data}, \bibinfo{address}{Newtown Square, PA}
  (\bibinfo{year}{1988}).

\bibitem[{\citenamefont{Fraundorf}(1981{\natexlab{a}})}]{Phil4}
\bibinfo{author}{\bibfnamefont{P.}~\bibnamefont{Fraundorf}},
  \bibinfo{journal}{Ultramicroscopy} \textbf{\bibinfo{volume}{7}},
  \bibinfo{pages}{203} (\bibinfo{year}{1981}{\natexlab{a}}).

\bibitem[{\citenamefont{Boisen and Gibbs}(1985)}]{Boisen}
\bibinfo{author}{\bibfnamefont{M.~B.} \bibnamefont{Boisen}} \bibnamefont{and}
  \bibinfo{author}{\bibfnamefont{G.~V.} \bibnamefont{Gibbs}},
  \emph{\bibinfo{title}{Mathematical crystallography}}
  (\bibinfo{publisher}{Mineralogical Society of America},
  \bibinfo{year}{1985}), vol.~\bibinfo{volume}{15} of
  \emph{\bibinfo{series}{Reviews in Mineralogy}}.

\bibitem[{\citenamefont{Spence and Zuo}(1992)}]{SpenceZuo}
\bibinfo{author}{\bibfnamefont{J.~C.~H.} \bibnamefont{Spence}}
  \bibnamefont{and} \bibinfo{author}{\bibfnamefont{J.~M.} \bibnamefont{Zuo}},
  \emph{\bibinfo{title}{Electron Microdiffraction}} (\bibinfo{publisher}{Plenum
  Press}, \bibinfo{year}{1992}), pp. \bibinfo{pages}{265--266}.

\bibitem[{\citenamefont{Fraundorf}(1987)}]{Phil27}
\bibinfo{author}{\bibfnamefont{P.}~\bibnamefont{Fraundorf}},
  \bibinfo{journal}{Ultramicroscopy} \textbf{\bibinfo{volume}{22}},
  \bibinfo{pages}{225} (\bibinfo{year}{1987}).

\bibitem[{\citenamefont{Spence}(1988)}]{Spen36}
\bibinfo{author}{\bibfnamefont{J.~C.~H.} \bibnamefont{Spence}},
  \emph{\bibinfo{title}{Experimental High-Resolution Electron Microscopy}}
  (\bibinfo{publisher}{Oxford University Press}, \bibinfo{address}{New York},
  \bibinfo{year}{1988}), pp. \bibinfo{pages}{87--89}, \bibinfo{edition}{2nd}
  ed.

\bibitem[{\citenamefont{Malm and O'Keefe}(1997)}]{Malm39}
\bibinfo{author}{\bibfnamefont{J.-O.} \bibnamefont{Malm}} \bibnamefont{and}
  \bibinfo{author}{\bibfnamefont{M.~A.} \bibnamefont{O'Keefe}},
  \bibinfo{journal}{Ultramicroscopy} \textbf{\bibinfo{volume}{68}},
  \bibinfo{pages}{13} (\bibinfo{year}{1997}).

\bibitem[{\citenamefont{Fraundorf}(1981{\natexlab{b}})}]{Phil7}
\bibinfo{author}{\bibfnamefont{P.}~\bibnamefont{Fraundorf}},
  \bibinfo{journal}{Ultramicroscopy} \textbf{\bibinfo{volume}{6}},
  \bibinfo{pages}{227} (\bibinfo{year}{1981}{\natexlab{b}}).

\bibitem[{\citenamefont{Selby}(1972)}]{Selb31}
\bibinfo{author}{\bibfnamefont{S.~M.} \bibnamefont{Selby}},
  \emph{\bibinfo{title}{Standard Mathematical Tables}} (\bibinfo{publisher}{The
  Chemical Rubber Co.}, \bibinfo{address}{Cleveland, Ohio},
  \bibinfo{year}{1972}), p. \bibinfo{pages}{199}, \bibinfo{edition}{twentieth}
  ed.

\bibitem[{\citenamefont{Liu et~al.}(1989)\citenamefont{Liu, Ming, and
  Hong}}]{Liu5}
\bibinfo{author}{\bibfnamefont{Q.}~\bibnamefont{Liu}},
  \bibinfo{author}{\bibfnamefont{Q.-C.} \bibnamefont{Ming}}, \bibnamefont{and}
  \bibinfo{author}{\bibfnamefont{B.}~\bibnamefont{Hong}},
  \bibinfo{journal}{Micron and Microscopea Acta}
  \textbf{\bibinfo{volume}{20}}(\bibinfo{number}{34}), \bibinfo{pages}{255}
  (\bibinfo{year}{1989}).

\bibitem[{\citenamefont{Tambuyser}(1985)}]{Tamb6}
\bibinfo{author}{\bibfnamefont{P.}~\bibnamefont{Tambuyser}},
  \bibinfo{journal}{Metallography} \textbf{\bibinfo{volume}{18}},
  \bibinfo{pages}{41} (\bibinfo{year}{1985}).

\bibitem[{\citenamefont{Qin et~al.}(1998)\citenamefont{Qin, Shi, Li, James,
  Siriwardane, and Fraundorf}}]{Qin30}
\bibinfo{author}{\bibfnamefont{W.}~\bibnamefont{Qin}},
  \bibinfo{author}{\bibfnamefont{W.}~\bibnamefont{Shi}},
  \bibinfo{author}{\bibfnamefont{J.}~\bibnamefont{Li}},
  \bibinfo{author}{\bibfnamefont{W.}~\bibnamefont{James}},
  \bibinfo{author}{\bibfnamefont{H.}~\bibnamefont{Siriwardane}},
  \bibnamefont{and}
  \bibinfo{author}{\bibfnamefont{P.}~\bibnamefont{Fraundorf}}, in
  \emph{\bibinfo{booktitle}{Proc. 1998 Spring Meeting}}
  (\bibinfo{organization}{Materials Research Society}, \bibinfo{year}{1998}),
  vol. \bibinfo{volume}{520}, pp. \bibinfo{pages}{217--222}.

\bibitem[{\citenamefont{Liu et~al.}(1992)\citenamefont{Liu, Hiang, and
  Yao}}]{Liu3}
\bibinfo{author}{\bibfnamefont{Q.}~\bibnamefont{Liu}},
  \bibinfo{author}{\bibfnamefont{X.}~\bibnamefont{Hiang}}, \bibnamefont{and}
  \bibinfo{author}{\bibfnamefont{M.}~\bibnamefont{Yao}},
  \bibinfo{journal}{Ultramicroscopy} \textbf{\bibinfo{volume}{41}},
  \bibinfo{pages}{317} (\bibinfo{year}{1992}).

\bibitem[{\citenamefont{Liu}(1990)}]{Liu1}
\bibinfo{author}{\bibfnamefont{Q.}~\bibnamefont{Liu}}, \bibinfo{journal}{Micron
  and Microscopea Acta} \textbf{\bibinfo{volume}{21}}(\bibinfo{number}{12}),
  \bibinfo{pages}{105} (\bibinfo{year}{1990}).

\bibitem[{\citenamefont{Williams and Carter}(1996)}]{Will37}
\bibinfo{author}{\bibfnamefont{D.~B.} \bibnamefont{Williams}} \bibnamefont{and}
  \bibinfo{author}{\bibfnamefont{C.~B.} \bibnamefont{Carter}},
  \emph{\bibinfo{title}{Transmission Electron Microscopy}}
  (\bibinfo{publisher}{Plenum Press}, \bibinfo{address}{New York},
  \bibinfo{year}{1996}), pp. \bibinfo{pages}{465--468}.

\bibitem[{\citenamefont{Qin and Fraundorf}(2000)}]{PaperC}
\bibinfo{author}{\bibfnamefont{W.}~\bibnamefont{Qin}} \bibnamefont{and}
  \bibinfo{author}{\bibfnamefont{P.}~\bibnamefont{Fraundorf}}, in
  \emph{\bibinfo{booktitle}{Proc. 58th Ann. Meeting}}
  (\bibinfo{organization}{Microscope Society of America},
  \bibinfo{year}{2000}), pp. \bibinfo{pages}{1038--1039}.

\bibitem[{\citenamefont{Fraundorf and Qin}(2001)}]{Maps}
\bibinfo{author}{\bibfnamefont{P.}~\bibnamefont{Fraundorf}} \bibnamefont{and}
  \bibinfo{author}{\bibfnamefont{W.}~\bibnamefont{Qin}}, in
  \emph{\bibinfo{booktitle}{Proc. 59th Ann. Meeting}}
  (\bibinfo{organization}{Microscope Society of America},
  \bibinfo{year}{2001}), pp. \bibinfo{pages}{?--?}

\bibitem[{\citenamefont{Treacy and Gibson}(1993)}]{Treacy1}
\bibinfo{author}{\bibfnamefont{M.~M.~J.} \bibnamefont{Treacy}}
  \bibnamefont{and} \bibinfo{author}{\bibfnamefont{J.~M.}
  \bibnamefont{Gibson}}, \bibinfo{journal}{Ultramicroscopy}
  \textbf{\bibinfo{volume}{52}}, \bibinfo{pages}{31} (\bibinfo{year}{1993}).

\bibitem[{\citenamefont{Treacy and Gibson}(1996)}]{Treacy2}
\bibinfo{author}{\bibfnamefont{M.~M.~J.} \bibnamefont{Treacy}}
  \bibnamefont{and} \bibinfo{author}{\bibfnamefont{J.~M.}
  \bibnamefont{Gibson}}, \bibinfo{journal}{Acta Cryst. A}
  \textbf{\bibinfo{volume}{52}}, \bibinfo{pages}{212} (\bibinfo{year}{1996}).

\bibitem[{\citenamefont{Voyles et~al.}(2000)\citenamefont{Voyles, Gibson, and
  Treacy}}]{Gibson1}
\bibinfo{author}{\bibfnamefont{P.~M.} \bibnamefont{Voyles}},
  \bibinfo{author}{\bibfnamefont{J.~M.} \bibnamefont{Gibson}},
  \bibnamefont{and} \bibinfo{author}{\bibfnamefont{M.~M.~J.}
  \bibnamefont{Treacy}}, \bibinfo{journal}{Journal of Electron Microscopy}
  \textbf{\bibinfo{volume}{49}}, \bibinfo{pages}{259} (\bibinfo{year}{2000}).

\bibitem[{\citenamefont{Gibson and Treacy}(1997)}]{Gibson2}
\bibinfo{author}{\bibfnamefont{J.~M.} \bibnamefont{Gibson}} \bibnamefont{and}
  \bibinfo{author}{\bibfnamefont{M.~M.~J.} \bibnamefont{Treacy}},
  \bibinfo{journal}{Phys. Rev. Lett.} \textbf{\bibinfo{volume}{78}},
  \bibinfo{pages}{1074} (\bibinfo{year}{1997}).

\bibitem[{\citenamefont{Gibson and Treacy}(1998)}]{Gibson3}
\bibinfo{author}{\bibfnamefont{J.~M.} \bibnamefont{Gibson}} \bibnamefont{and}
  \bibinfo{author}{\bibfnamefont{M.~M.~J.} \bibnamefont{Treacy}},
  \bibinfo{journal}{J. Non-Cryst. Solids} \textbf{\bibinfo{volume}{231}},
  \bibinfo{pages}{99} (\bibinfo{year}{1998}).

\end{thebibliography}


\end{document}